# Maximum Match Subsequence Alignment Algorithm – Finely Grained(MMSAA – FG)

Bharath Reddy
Process Automation R&D
Schneider-Electric
Lake Forest, U.S.A
Bharath.reddy@se.com

Richard Fields
Process Automation R&D
Schneider-Electric
Lake Forest, U.S.A
Richard.fields@se.com

*Abstract*—Sequence alignment is common nowadays as it is used in many fields to determine how closely two sequences are related and at times to see how little they differ. In computational biology / Bioinformatics, there are many algorithms developed over the course of time to not only align two sequences quickly but also get good laboratory results from these alignments. The first algorithms developed were based of a technique called Dynamic Programming, which were very slow but were optimal when it comes to sensitivity.

To improve speed, more algorithms today are based of heuristic approach, by sacrificing sensitivity. In this paper, we are going to improve on a heuristic algorithm called MASAA (Multiple Anchor Staged Local Sequence Alignment Algorithm) – and MASAA -S which we published previously. This new algorithm appropriately called MMSAA – FG, stands for Maximum Match Subsequence Alignment Algorithm – Finely Grained. The algorithm is based on suffix tree data structure like our previous algorithms, but to improve sensitivity, we employ adaptive seeds, and finely grained perfect match seeds in between the already identified anchors. We tested this algorithm on a randomly generated sequences, and Rosetta dataset where the sequence length ranged up to 500 thousand.

## I. Sequence Alignment

In computational biology or Bioinformatics, a sequence is either an RNA, DNA or a protein string made up of their representative character set. DNA ( A, C, G, T) , RNA ( A, C, G, U) and protein molecules (A, R, N, D, C, Q, E, G, H, I, L, K, M, F, P, S, T, W, Y, V) can be re represented as strings of letters from their alphabet set [1] [2] [3] [25].

A sequence alignment is a way of arranging these sequences made of their representative characters with an objective to find the regions of '*similarity*'. These similarities would then provide additional information on the functional, structural, evolutionary and other interest between the sequences in study. Aligned sequences are represented in rows, stacked up one on top of the other as shown below in Figure 1.

```
T T A T A G A G G _ A C A _ C G
| | | |   | |   | | |   | |
_ _ A T A G _ G G G A C A T G G
```

Figure 1. Example of a sequence alignment. [25]

In fig 1, there are regions, where the two sequences are aligned perfectly, these regions are called 'similar region. In some regions, special characters such as '-', also known as indels are present. These indels represents a mutation (change) or could be looked at as deletion from the other sequence's point of view [25].

Pairwise sequence alignment is used to find conserved regions in two sequences. Multiple sequence alignments are used to find common regions in more than 2 sequences. Pairwise Sequence alignment [25]. Pairwise sequence alignment is a first step and represents at times the first step in many bioinformatics solutions. In many multiple sequence alignment algorithms this remains the first step, especially the linear multiple sequence alignment algorithms.

Pairwise sequencing as in Fig 1, is alignment between 2 sequences of the same kind, DNA, RNA or Protein. The alignment would then throw some knowledge on the divergence of one sequence over the other in some cases or similarity in some cases.

Pairwise sequence alignment can be classified into local sequence alignment and global sequence alignment. Local sequence alignment finds the best approximate subsequence match within two given sequences while the global sequence alignment takes the entire sequence into consideration [25].

Local sequence alignments are therefore designed to search subregions within the two sequences. For finding similar (biologically conserved) regions, which may or may not be preserved in order or orientation, local sequence alignment is very useful. It is typically used to find similarity between two divergent sequences and for fast database searches for similar sequences [25]. Since, it is trying to find subregions and not the sequence in its entirety, local sequence alignment usually takes less computation time when compared to global sequence alignment algorithms [25].

Some of the most popular local sequence algorithms are Smith-Waterman [4], FASTA [5], BLAST [6], GappedBLAST[7], BLASTZ[8], PatternHunter[9], YASS[10], LAMBDA[11], USearch [12], LAST [13], and ALLAlign [15].

Popular global sequence alignment algorithms including Optimal, and Heuristic based are Needleman and Wunsch [18], GLASS [19], WABA [20], AVID [21], AlignMe [14] and GLASS [22]. We will not be discussing global alignment algorithms in this paper.

The alignment algorithms performances in the literature are based off several key measurements. This is very touchy, subjective topic and can vary from algorithm to algorithm. Some are based on type of sequence (they can be only for DNA, or RNA), length (some algorithms can be good for shorter than others), Measure of accuracy (since there is no standard here, this is a controversial and subjective), Speed of alignment (time taken to align the sequences in study) and lastly memory efficiency (how much memory is taken to find the alignment) [25]. Today, AI is being used to align pairwise sequence alignment, and in other fields [26], but is not in the scope of this paper.

## II. LITERATURE

In this section we will talk about the popular algorithms in local sequence alignment. Our algorithm is greatly influenced by previous algorithms and has taken clues and has questioned at times on the approach and ideas of the previous algorithms. We first start the section with optimal and then heuristic algorithms.

### A. Local Sequence Alignment algorithms

Smith-Waterman Algorithm is an optimal local sequence alignment algorithm, employing a technique called '*Dynamic Programming*', where a problem is broken into smaller problems and solving these smaller problems recursively [25]. The solutions to the subproblems are saved and are brought together to find the solution to the entire problem.

This optimal algorithm produces an optimal local sequence alignment between two sequences S1 and S2 of length m and n, respectively, in time and space equal to $O(mn)$ [25]. As the sequence length increase the performance decreases exponentially.

To overcome short comings of the optimal algorithm, Heuristic algorithms were developed later. Heuristic algorithms find a near-optimal solutions sacrificing little sensitivity for speed. Since it is near perfect and can calculate long sequences running in billions of characters, heuristic algorithms are preferred. All heuristic algorithms run in stages, stages where they find the maximum subsequence and try to find the regions arounds them to get the sequences local sequence alignment. These subsequences are called seeds or anchors depending on whether the subsequence is one large word (anchor) or small words made of few characters (seed).

The first heuristic algorithm we are going to talk is FASTA, which stands for FAST-ALL is a heuristic algorithm developed by Lipman and Pearson [5] [25]. FASTA uses a look-up table to find perfect subsequence matches of size 'l' and then hashing them. Time is saved for searching seeds of length 'l', then proceeds to find such seeds in a diagonal path, since the final alignment is more likely to be found in this diagonal length. It then uses a directed weighted graph for regions in between the seeds along the best diagonal found to stitch the seeds. The advantage of FASTA over Optimal algorithm is the speed but if there are 2 optimal diagonals or if the seeds are smaller than the size deployed in FASTA, then the algorithm loses considerable sensitivity [25].

Basic Local Alignment Search Tool BLAST [6], uses a look-up table to identify seeds and is faster than FASTA [25]. Sliding window technique is employed to find all good neighbors seeds for each seed it finds in both directions [25]. When all seeds are found, it then proceeds to find the seeds (HSP, high scoring pairs), and extend them until they fall under a threshold score 'k'. These HSP are then stitched using a restricted dynamic programming which is a version of Smith-Waterman Algorithm [25].

BLAT – BLAST like alignment tool [17] is much faster than BLAST. BLAT differs from previous algorithms in the way sequences are indexed. "non-overlapping seeds of S2 are run through the database of sequence and then a new scan is run linearly through the S1, whereas BLAST builds an index of S1 and then scans linearly through the database" [17]. This saves time. After this stage, it then searches for seeds with some mismatches 'n' in them around the seeds it found earlier. Then the HSPs of seeds and mismatch seeds are extended like BLAST to form a final alignment. As with BLAST, BLAT cannot find smaller homologous regions as the seeds taken as not small enough.

BLASTZ [8], is the fastest among the BLAST family of algorithms, employs a different method. All repeats in the sequence are removed [25]. It then looks for seeds of length 'l' with almost one-character transition. All seeds are then extended on both sides. For regions in between the seeds, it employs smaller seeds and uses optimal alignment to stitch these seeds to form final alignment. Since matched or repeat seeds are not used again, and transition seeds set to almost one, there is a possibility that this algorithm performs poorly when it is used for divergent sequences. Meaning, say a drosophila and a pig DNA.

PatternHunter [9] introduced a seed called spaced seed to further improve the sensitivity and speed. It uses a combination of priority queues variation of red-black tree, queue and hash table to achieve speed [9]. A spaced seed is a generic seed which is converted from A, C, G, T to Binary form e.g.: 1010101,4 where 1 is a match and 4 is a score [25]. It then finds the best diagonal as in FASTA to find the final alignment [25]. The algorithm is written in JAVA, and encounters memory problems for long sequences.

UBlast [12] introduces a new technique by finding fewer good hits. Meaning, subsequences which are found least but are long, in order to improve speed on BLAST and MEGABLAST [16] which is algorithm from BLAST family. The technique is targeting more speed than sensitivity.

LAST [13] is recent algorithm. It uses adaptive alignment seeds; these adaptive seeds vary in length and the number of indels in them. So adaptive seeds can be of different lengths and weight. By weight, a score associated with the seed. The rest of the algorithm is very similar to BLAST. ALLAlign [15] is a new algorithm developed, however literature of this AWS based web algorithm is very limited.

LAMBDA [11] is new algorithm for protein sequence alignment. It implements a technique where there are more than 1 protein sequences as the target sequences to be aligned with a pre indexed database set of all other know sequences [25]. It is optimized for big or large biological data and uses a Suffix tree to get the maximal common subsequences or maximal unique sequences as our algorithm in this paper, amongst them and then goes about aligning these subsequences against a pre indexed database (pre indexed based off suffix array) [25].

MASAA [1] [3] introduced in 2008 is based on Ukkonen suffix [25] tree. The algorithm uses double indexing and back tracking and identifies maximum match subsequences (MMSS) [25]. In the subsequent stages, it finds perfect and near perfect seeds and stitches the local alignment in the last stages.

MASAA – S [25] was introduced in 2019 which is similar to MASAA but uses adaptive seeds in between the MMSS, In the later stages it uses perfect seeds to improve sensitivity. The algorithm is also more sensitive than MASAA but comparable in speed to MASAA. We use this to question two things, if adaptive seeds are not incurring speed penalty and are improving the sensitivity, then how far can we further the sensitivity without sacrificing the speed. The algorithm in this paper addresses these questions.

### III. MOTIVATION

The motivation for this paper is our previous paper MASAA – S which is based on the same technique using different seed structure. That paper questioned how far we can go in terms of sensitivity without sacrificing the sensitivity. This paper further enhances the sensitivity, but we think we are hit a threshold when it comes to sensitivity and pushed the boundary using this technique to boundary. In this paper, we introduce an algorithm technique which extends MASAA and MASAA – S [1], [3], [25] an algorithm we introduced in 2008, 2019 by making it more sensitive and relatively faster at the same time compared to others.

### IV. MMSAA – S (MAXIMUM MATCH SUBSEQUENCE ALIGNMENT ALGORITHM – FINELY GRAINED)

MMSAA – FG is like MASAA and MASAA - S in the first two of five stages, but it completely differs in the kind of seeds selected in between the maximum match subsequences (MMSS), extension anchor stage and final stage - stitching the whole alignment. We will explain in detail in the coming sections.

#### A. Finding MMSSs

In this stage two sequence are merged into one big sequence by introducing a special character in between the two sequences. A suffix tree is built for this big sequence. MMSS whose length 'l' is greater than the arbitrary length. The arbitrary length is set at 1/3rd the length of the longest MMSS. Using Ukkonen version of suffix tree with pointers, we employ back tracking are finds all MMSSs between the sequences. Since these are MMSS and not seeds, there is no question of noise in this selection.

#### B. MMSS anchors within a 60% of the length

All non-overlapping and non-crossing are chosen as in MMSAA -FG [1][3] [25]. In this step we want to select all anchors which fall in our neighborhood. A good neighbor is distance from the previous MMSS within which we must find the next MMSS. We always start from the longest subsequence we found in the previous stage. We try to find the MMSS with in 60% of the length of that, and then move the next MMSS, then find the MMSS which is inside 60% of that MMSS and so on, this is different from previous MASAA step.

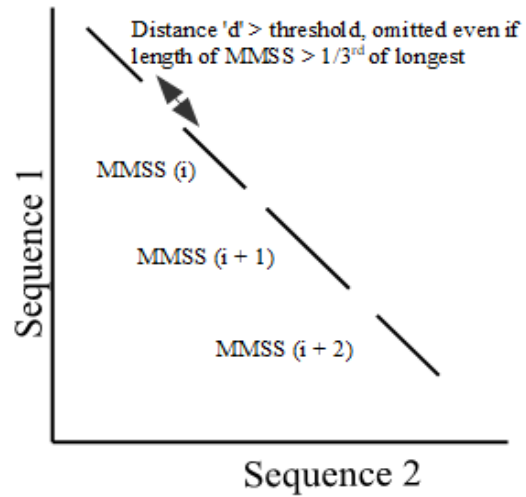

Fig 2: MMSS in a good neighborhood are considered [25]

#### C. Finding adaptive seeds In Between MMSSs

This step can be broken down into 2 steps.
  I. In this step, we find all the adaptive seeds which are found using the Suffix tree. These adaptive seeds are set at a size 20 with 6 mismatches. This step also finds more seeds while keeping the sensitivity. This also aids in speed.
  II. In this step, we find perfect match seeds of length 4 and then 2 after that. between the adaptive seeds found in the previous step. These seeds should be within 1/3rd the distance from the adaptive seeds found. The size of seed is fixed in the algorithm primarily because, a K-mer of size between 8 or less is more likely to find more seeds than a K-mer of size ranging from 12 - 48 [16] [13] where the authors have clearly established that there is a great propensity to get more hits when the size is between 8 – 12 [25]. Both the steps are shown in Fig 3.

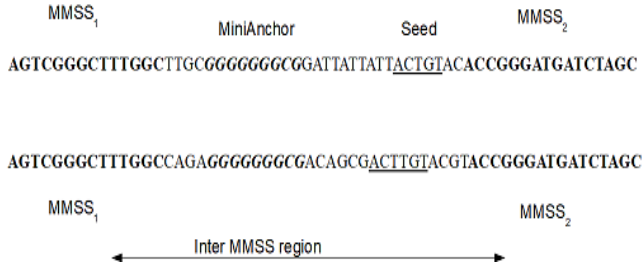

Fig 3: Inter MMSS, Mini Anchor and small seed [25]

### D. Finding the non criss crossing adaptive and perfect seeds

This algorithm now finds all overlapping and non-crossing matching from the adaptive seeds and perfect seed selection in the previous step. To identify anchors from overlapping and crossing matches, we use the heuristic of "closeness. The question of overlapping does not arise in the case of adaptive seeds because all overlapping seeds would amount to a bigger seed which would have been picked up already using the suffix tree in the previous step part 1. The heuristic of closeness would arise only in the case of selecting perfect seeds of length 4 and 2 either in between MMSSs and Mini-Anchors or in between MMSSs. Hence this is intensive and relatively time-consuming part when compared to our previous algorithms.

### E. Final Stitching

The final stitching is like our previous algorithm The MMSSs anchors and Mini-Anchors found in previous step, forms most of the final alignment [25]. The anchors are extended on both sides starting from left to right. The algorithm terminates when all MMSS are extended and included in the final alignment.

## V.  IMPLEMENTATION

The algorithm is implemented in C language and the core part of the program is the Ukkonen Suffix tree with pointers. This suffix tree gives the ability to back track to find all MMSSs without sacrificing speed. It is at the same time simple and robust. The adaptive seeds are also found out the same way in between the MMSS and only their start and end positions are noted.
. The algorithm is pictorially shown in figure.

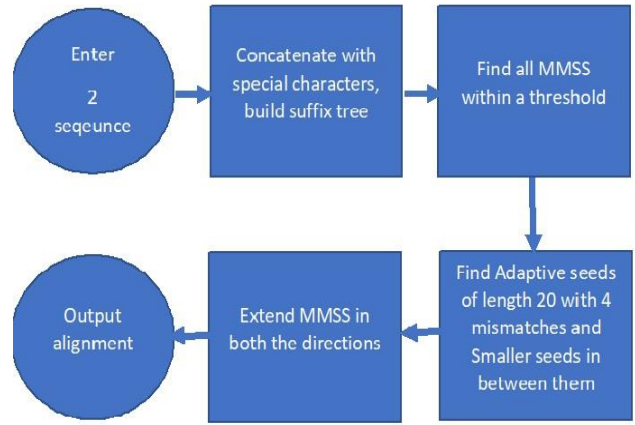

Fig 4: Algorithm

## VI.  SPEED, SENSITIVITY AND EXPERIMENTATION

The metrics we subject this algorithm is speed and sensitivity. Although Speed is straight forward, sensitivity is subjective from our experience and literature. There is no standard for sensitvity. Different programs use sensitivity metrics, which can influence the results [4]. Some use 'correctly aligned residue pairs' divided by number of residue pairs in the reference sequence. Others, Total column score (TCS), the number of correctly aligned columns divided by the number of columns [25].

In this algorithm we use fraction of Exon length aligned to the corresponding exon as our sensitivity. Like in MASAA and MASAA -S we have randomly generated sequences up to 500 thousand in length. To check the sensitivity of the algorithm, we used the same ROSETTA dataset [3][23] for homologous sequences and used our own set of different genes from different animals to compare our algorithms with the rest of the algorithms

## VII.  EXPERIEMNTAL ANALYSIS

In the experimental results, we randomly generated sequences whose length is from 100k to 500k and compared the speed of alignment. For smaller sequences the speed is much faster than BLASTZ and compares well with our previous algorithms. However, when the sequences grow, then the speed of our new algorithm is slower than that previous algorithm and almost touching BLASTZ territory. We believe it will perform slower as the length of the sequence increases further. We did not compare the BLASTZ, MASAA-S and MMSAA - FG with LAMBDA, AllAlign for reason, because it needed an index database first and for a randomly generated sequences it was challenging [25]. For ALLAlign, we could not find source code to download and compare, checking the performance of the sequence on a server was not clinical. Although we know LAMBDA is 500x faster than BLASTZ, here we assume that LAMBDA is faster than MMSAA-FG too.

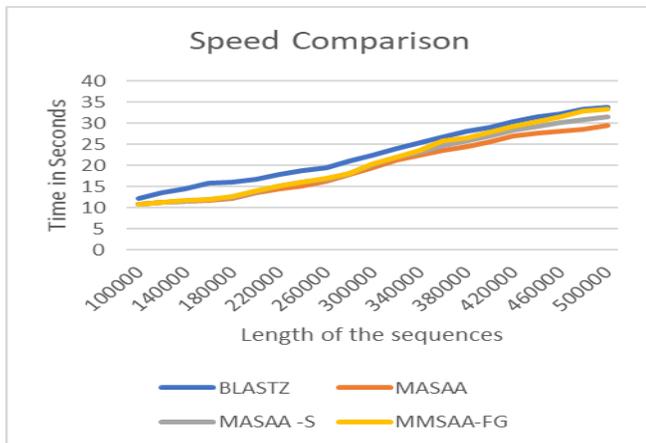

Fig 5: Speed Comparison on different sequence length

For sensitivity, we compared the exon coverage from all four algorithms. We compared the performance of four algorithms on the Rosetta Dataset. Table 1 shows the percentage of exon coverage, here we think MMSAA-FG performs better than MASAA and MASAA-S, we attribute this to smaller anchors seeds being chosen once the Large Anchors are already chosen in the first step. Also, smaller seeds of different length <= 4 plays a role. This step is little performance heavy (speed) as we see in the fig 5.

TABLE I. SENSITIVITY COMPARISION ON ROSETTA DATASET

| Algorithm | % of exon coverage | | |
|---|---|---|---|
| | 100 exon | 90 exon | 70 exon |
| BLASTZ | 94 | 97 | 98 |
| MASAA | 94 | 94 | 96 |
| MASAA - S | 94 | 96 | 97 |
| MMSAA - FG | 94 | 96 | 98 |

MMSAA-FG performed better than BLASTZ and our previous algorithms on divergent sequences, so, we compared the four algorithms for divergent sequences on different genes from ROSETTA dataset and compared the percentage of exon coverage. MMSAA-FG performed better than MASAA, MASAA-S and BLASTZ. We attribute this performance to Mini Anchor and perfect match seeds of smaller size in the inter anchor and inter MINI anchors. This is shown in Fig 6.

## VIII. CONCLUSION

In this paper, we have proposed a new algorithm which is not only faster than BLAZTZ for smaller sequences but more sensitivity than BLAZTZ and our previous algorithms on divergent sequences. On Rosetta dataset, we found that the algorithm performs closer to BLASTZ. We attribute this to smaller seeds in between the MMSS and mini anchors. In the future, we would want to extend the algorithm with a different stitching algorithm which is faster than ours and perhaps parallelize.

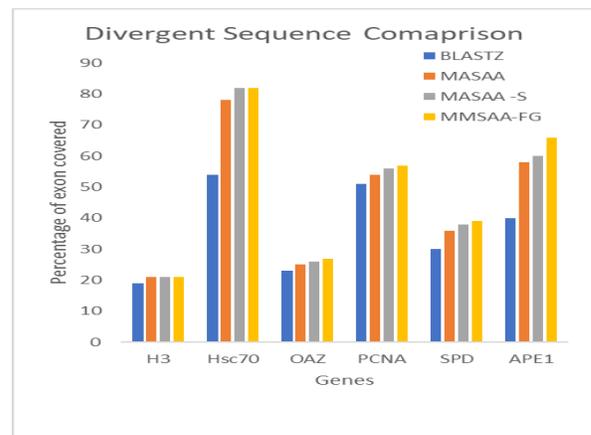

Fig 6: Exon coverage in divergent sequences

| Full Name | Email address | Position | Research Interests | Personal website (if any) |
|---|---|---|---|---|
| Mr. Bharath Reddy | Bharath.reddy@se.com | Senior Software Designer | Bioinformatics Mutual Exclusion and AI | n/a |
| Mr. Richard Field | Richard.fields@se.com | reviewer | Bioinformatics | n/a |